\documentclass{article}

\usepackage[left=2cm,top=1.8cm,bottom=1.5cm,right=2cm,headsep=.45cm,footskip=.8cm]{geometry} % set the margins
\usepackage{amsmath}
\usepackage{graphicx}
\usepackage[colorlinks=true, allcolors=blue]{hyperref}

\usepackage{leading}
\leading{12.65pt} % To get equivalent of word's single spaced text. The default leading in word is I think 1.15 x 11pt = 12.65pt. 

\usepackage[svgnames]{xcolor}
\usepackage{versions}

\includeversion{b1}\excludeversion{b2}

\usepackage{fancyhdr}
\RequirePackage[explicit]{titlesec} 
\titleformat{\part}
  [display] % Display style
  {\filcenter\LARGE\bfseries\color{white}} % Format for the section title
  {} % Label
  {0pt} % Separation between label and title
  {\vspace{-10mm}\colorbox{RoyalBlue}{\parbox{\dimexpr\textwidth-2\fboxsep}{\center #1\vspace{4mm}}}} % Code before the title
  [] % Code after the title
\usepackage{etoolbox}
\renewcommand{\refname}{}
\patchcmd{\thebibliography}{\section*{\refname}}{}{}{}

\usepackage[linewidth=1pt]{mdframed} % for abstract in B1

\usepackage{enumitem}

\usepackage{tabularx} % for Funding ID
\usepackage{colortbl}% for Funding ID

\hypersetup{
    colorlinks=true,
    linkcolor=blue,
    filecolor=magenta,      
    urlcolor=red,
    citecolor=DarkGreen,
    pdftitle={Overleaf Example},
    pdfpagemode=FullScreen,
    }
\usepackage[T1]{fontenc}    % proper output font encoding
\usepackage[utf8]{inputenc} % read source as UTF‑8
\usepackage{lmodern}  
\usepackage{siunitx}   
\usepackage{eurosym}
\usepackage[most]{tcolorbox}

\usepackage{caption}
\usepackage{adjmulticol}
\let\OLDthebibliography\thebibliography
\renewcommand\thebibliography[1]{%
  \OLDthebibliography{#1}%
  \setlength{\parskip}{0pt}%
  \setlength{\itemsep}{2pt plus 0.3ex}%
}

\newcommand{\HII}{\mbox{H\,{\sc ii}}}

\newcommand{\AV}{\mbox{$A_{5495}$}}

\newcommand{\Msol}{\mbox{M$_\odot$}}

\title{Why the northern hemisphere needs a \\ 30–40 m telescope and the science at stake: \\ Massive stars in spiral galaxies}
\author{Jesús Maíz Apellániz$^1$, 
        Sergio Simón-Díaz$^{2,3}$,
        Artemio Herrero$^{2,3}$,
        Sara R. Berlanas$^1$, \\
        J. Miguel Mas Hesse$^1$,
        Ignacio Negueruela$^4$,
        Gonzalo Holgado$^{2,3}$, and
        Miriam Garcia$^5$,
        \\
        \\
        $^1$ Centro de Astrobiología, campus ESAC. \num{28692} Villanueva de la Cañada, Madrid, Spain. \\
        $^2$ Instituto de Astrofísica de Canarias. \num{38200} La Laguna, Tenerife, Spain. \\
        $^3$ Universidad de La Laguna. \num{38206} La Laguna, Tenerife, Spain. \\
        $^4$ Universidad de Alicante. 03\,690 San Vicente del Raspeig, Alicante, Spain. \\
        $^5$ Centro de Astrobiología, campus Torrejón. \num{28850} Torrejón de Ardoz, Madrid, Spain. 
        }

\begin{document}

\maketitle

\begin{abstract}
This document discusses the three main lines expected to dominate massive-star research in the 2040s, namely:

\begin{enumerate}[nosep]
 \item[\textbf{(1)}] The role of metallicity in stellar evolution, especially in determining the end products such as gravitational-wave progenitors.
 \item[\textbf{(2)}] The initial mass function from the most massive stars to substellar objects.
 \item[\textbf{(3)}] The role of the environment in the different modes of star formation from compact star clusters to born-this-way associations and from massive clusters to small stellar groups.
\end{enumerate}

More specifically, we present the contributions to such science that would be enabled by a 30~m type telescope in the northern hemisphere studying spiral galaxies. Those can be grouped in three: our own Galaxy, the Milky Way; the other two spiral galaxies in the Local Group, M31 and M33; and other galaxies within 25 Mpc, such as M101, M51, and NGC~6946. This work is based on the fact that, as of today, no construction of a 30~m telescope has yet started in the northern hemisphere, so even in the best case scenario of such a hypothetical telescope, its full operation would not start until the late 2030s or early 2040s. It makes no assumptions about its location but supposes an instrumentation development similar to that of ELT.
\end{abstract}

\newpage
% HARMONI: Intermediate-resolution IFU spectroscopy in 0.47-2.45 microns
% MICADO: NIR Multi-AO imaging + spectroscopy
% METIS: MIR imaging + spectroscopy
% ANDES: High-resolution (100 000) spectroscopy in 0.40-1.80 microns, single object + IFU
% MOSAIC: Nulti-object spectroscopy in 0.39-1.80 microns

\section{Introduction}

$\,\!$\indent Massive stars are the great galactic influencers as, solar mass for solar mass, no other astronomical objects exert as many effects in galactic evolution in the form of ionising radiation, chemical enrichment, and injection of kinetic energy into the surrounding stars and ISM. Yet, they are scarce and often located in dust-enshrouded environments, which often makes massive-star astronomy a game of picking the right location to study its targets. Under those circumstances, in this white paper we first present the main lines of research expected to be significant for massive stars in the 2040s and then we point the locations in spiral galaxies that are unique to the northern hemisphere and not accessible from the southern hemisphere. Those are divided in three blocks: the Milky Way (with the Cygnus sightline as the most inetresting location); M31 and M33 in the Local Group; and other northern spiral galaxies such as M51, M101, and NGC~6946.

\section{Main research lines for massive stars in the 2040s}

$\,\!$\indent The main lines of research that we expect to be significant for massive stars in the 2040s are:

\begin{enumerate}[nosep]
 \item[\textbf{(1)}] The role of metallicity $Z$ -- ranging from near-primordial to supersolar -- in stellar evolution. Metallicity impacts the evolution of massive stars primarily through mass loss, which is higher at large values of $Z$. This, in turn determines the nature of the final end products,  which we are observing as gravitational-wave (GW) progenitors. Nearby spiral galaxies are an excellent laboratory to explore this issue, as a 30~m telescope offers the possibility of spatially resolved analyses of individual stars across their metallicity gradients.
 \item[\textbf{(2)}] The initial mass function (IMF) from the most massive stars to substellar objects. At near-primordial metallicities the IMF is expected to be top heavy, with a significant number of very massive stars (VMSs, with masses above 120~\Msol), but such stars are also observed in galaxies such as the Large Magellanic Cloud (LMC).     
 \item[\textbf{(3)}] The role of the environment in the different modes of star formation (SF), from compact star clusters to born-this-way associations and from massive clusters to small stellar groups. Is the IMF constant in different environments? How soon is the present-day mass function (PDMF) observed in clusters altered by dynamical events such as stellar ejections? How does massive-star feedback into the ISM alter galactic structure and future star formation? 
\end{enumerate}

\section{The Cygnus sightline in the Milky Way} 

%\section*{Massive stars: the great galactic influencers}
%\subsection*{WP 1: Sample generation}

\begin{figure}[h!]
 \newlength{\imgw}
 \settowidth{\imgw}{\includegraphics{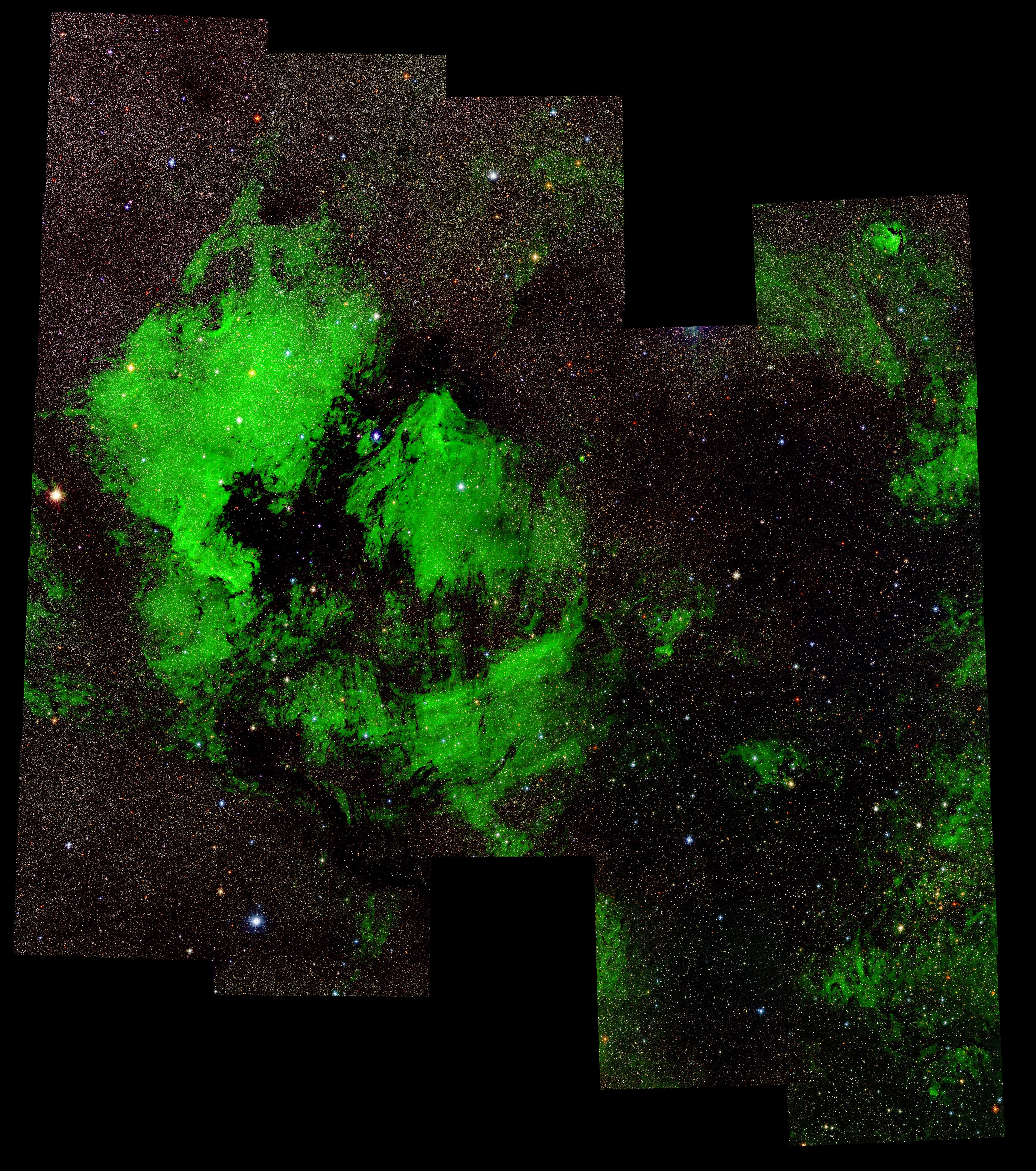}}
 \newlength{\imgh}
 \settoheight{\imgh}{\includegraphics{Cygnus_E_jplus1_subback_reduced.jpg}}
 \centerline{\includegraphics[trim=0.05\imgw{} 0.1\imgh{} 0.05\imgw{} 0.2\imgh{},clip,width=\linewidth]{Cygnus_E_jplus1_subback_reduced.jpg}}
% 44 000 x 50 200 px -> 39 600 x 35140 px -> 6.05 x 5.37 deg
 \caption{6.05$\times$5.37 sq.dg. (180~pc~$\times$~159~pc at a distance of 1.7~kpc) portion of the Cygnus sightline from the GALANTE survey 
          \cite{Maizetal21d} in F861M (R), F660N (G), and F515N (B).  The North America and Pelican nebulae are in the upper left ($d \sim$800~pc) and 
          the massive Cyg~OB2 region is in the lower right  ($d \sim$1.7~kpc, both in the Local spiral arm), where we also find more distant highly 
          extinguished objects such as Cyg~X-3 ($d \sim$7.4~kpc). Extinction is highly variable, with cases of $\AV >10$~mag even at distances less than 
          1~kpc. North is up and East is left.}
 \label{Cygnus}
\end{figure}

$\,\!$\indent The sightline in the direction of Cygnus is not accessible from most of the southern hemisphere and is unique in the sky because it is a 
MW spiral-arm tangent that contains: (a) the site of most
intense recent SF within 2~kpc (Cyg OB2); (b) the most-massive O-type binary within 1~kpc (Bajamar star); (c) other massive-star forming
regions in the MW Local Arm; and (d) more distant objects located in other MW spiral arms such as Cyg~X-3. Despite the proximity, the location at a tangent
makes extinction exceptionally large and requires large telescopes (even 30~m-class ones) to acquire high-quality data in the blue-violet region of the
spectrum, where the information content for OB stars is by far the highest. As examples, three stars in Fig.~\ref{Cygnus}, Bajamar star, Cyg OB2-12, and 2MASS J20395358+4222505,
would be among the brightest in the northern hemisphere if it were not for extinction.  Instead they are hard to locate in that image because they have
$\AV > 10$~mag \cite{Maizetal21a,Herretal22}.

Optical multi-object spectroscopy with a 30~m telescope would allow to fully characterize the massive- and intermediate-mass stellar population in the North America nebula, Cyg~OB2 and elsewhere along the sightline within 2~kpc, even for the targets with the highest extinctions. Equivalent NIR multi-object spectroscopy would allow to do the same with the low-mass and substellar pre-main sequence (PMS) populations. This comprehensive characterization would lead to the derivation of a robust IMF across the full mass spectrum and to explore the role of environmental conditions and stellar feedback in shaping the IMF and the dynamical evolution of young massive clusters at near-solar metallicities. The proximity of the analysed regions would allow for the inclusion of the full effects of binarity \cite{Maizetal19b} and ejected stars \cite{Maizetal22b} and from there determine the possible existence of variations from the canonical Kroupa IMF \cite{Krou02}.

Other Galactic sightlines should provide complementary information on the IMF. A 30~m northern telescope should also be able to peer into the very young W3 region (the Fish Head Nebula) and to analyze the IMF for the full mass spectrum (including brown dwarfs) in the outer regions of the MW, with metallicities expected to approach those of the SMC but at an order of magnitude in distance closer.

\section{M31 and M33: the Rosetta stones for spiral galaxies} 

$\,\!$\indent  M31 and M33 are the two other spiral galaxies in the Local Group besides the MW, which makes them the Rosetta stones for the study of spiral galaxies. They are only accessible from the northern hemisphere and, at the distance of M31 and M33, a 30-meter telescope should be able to resolve compact massive clusters for massive and intermediate-mass stars (Table~5 in \cite{Skidetal15} and Fig.~\ref{M33}).

\subsection{M31}

$\,\!$\indent M31 is the largest member of the Local Group and the closest spiral, at a distance of 790~kpc. Despite its proximity, our knowledge of its chemical properties is more imprecise in comparison with M33 or even more distant galaxies such as M101 due to its high inclination (77$^{\rm o}$), making the extinction stronger than in face-on galaxies. Bright, large \HII\ regions are scarce and, additionally, those observed have usually low excitation (hence making more difficult the determination of abundances via direct methods). There are few studies using \HII\ regions since the pioneering works by \cite{1981AJ.....86..989D,1982ApJ...254...50B}. More recently, \cite{1999AJ....118.2775G} determined a metallicity gradient of $-0.06$~dex/kpc for 20 \HII\ regions but based on the indirect method. The only analyses using massive stars (blue supergiants, or BSGs) are those by \cite{2000ApJ...541..610V,2001MNRAS.325..257S,2002A&A...395..519T} do not show evidence of a significant gradient but they only have a total of seven stars and significant scatter due to low number statistics: a larger sample is clearly needed, and this could be obtained with 10~m class telescopes in the next decade, as the PHAT survey done at HST resolution \cite{Dalcetal12} allows us to pick isolated BSGs with relative ease. 
However, a 30-m telescope incorporating mid- and high-resolution nIR spectroscopy (including both single object and IFU) will be needed to achieve, for the first time, a full characterization of the massive star population in M31, leveraging the limitations imposed by extinction. Eventually, this will provide unique datasets to perform investigations of the IMF of clusters and associations as a function of metallicity with emphasis on spatial resolution and a larger mass range compared to M51/M101

\subsection{M33}

\begin{figure}[t]
 \centerline{\includegraphics[width=\linewidth]{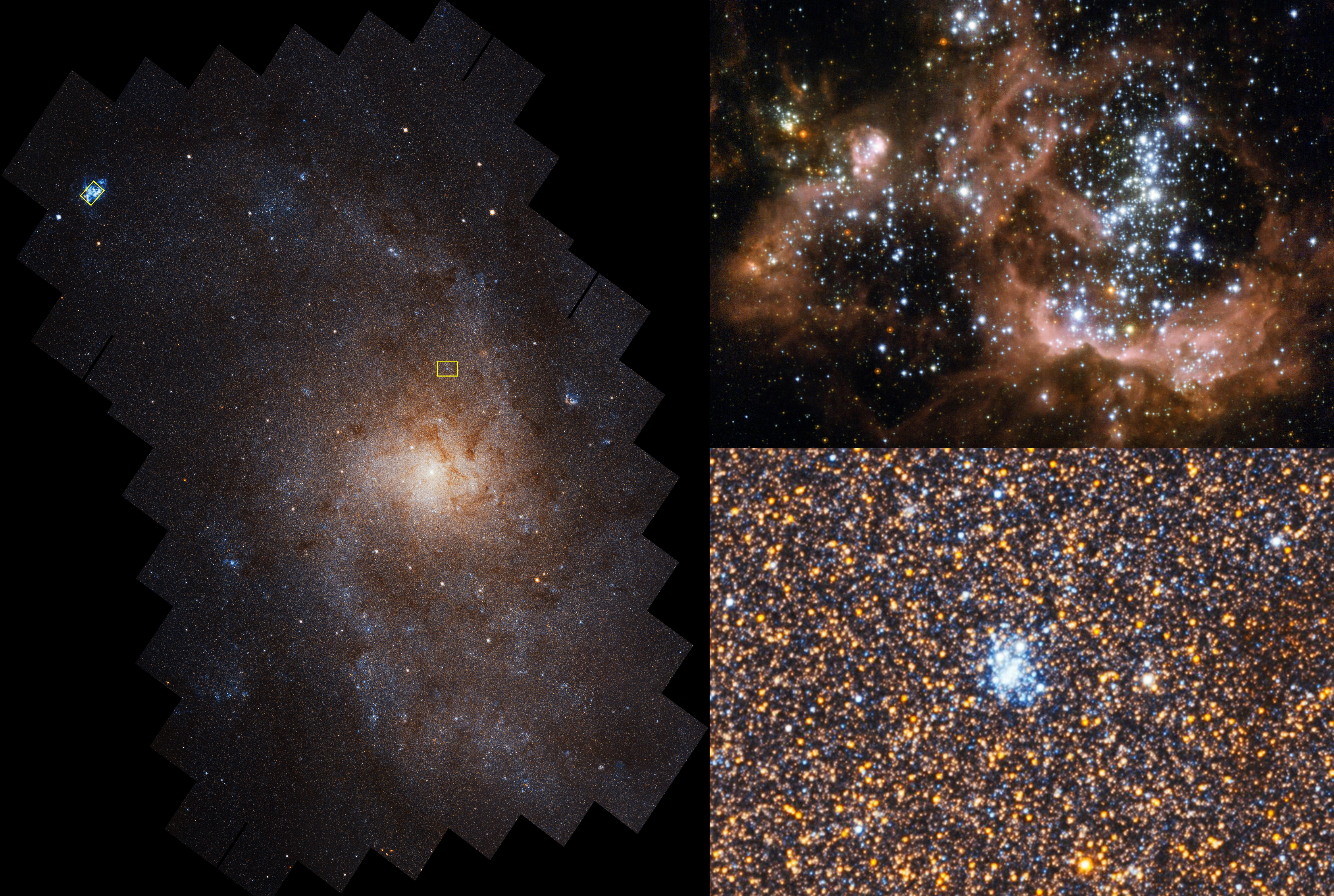}}
 \caption{(left) $18.8'\times 23.8'$ (4.60~kpc~$\times$~5.82~kpc) HST/ACS-WFC M33 mosaic from the PHATTER project \cite{Willetal21} with F435W used for 
          the cyan channel and F814W for the orange channel. North is up and East is left and the yellow rectangles correspond to the two 
          $31.35''\times 22.50''$ (128~pc~$\times$~92~pc) regions on the right panels. 
          (top right) HST/ACS-HRC image of NGC 604, the scaled OB association that is the largest site of recent massive-star formation in M33.
          The colors are from a combination of UV+optical filters (see \url{https://esahubble.org/images/potw1019a/}) obtained from HST program GO~9419
          (P.I.: Rodolfo Barbá). Given the intensity of SF and the location in the outer region of M33, most stars seen in the image are
          young and members of NGC~604. (bottom right) Zoom-in of the left panel into a young stellar cluster in the inner region of M33. The population
          in this case is dominated by the field, with a mixture of masses and ages and most of the light originating in low- and intermediate-mass red 
          giants. Given the compactness of the cluster, it is only partially resolved even with HST but could be easily resolved by a 30~m telescope 
          with adaptive optics.}
 \label{M33}
\end{figure}

$\,\!$\indent M33 is the third largest member of the Local Group of galaxies after M31 and the MW. It was the first of the galaxies in which a radial abundance gradient was recognised \cite{1971ApJ...168..327S}. Thanks to its modest distance (810 kpc) and almost face-on orientation (Fig.~\ref{M33}), it is the spiral galaxy with the largest amount of \HII\ data to study the distribution of abundances, using either direct or indirect methods \cite{2008ApJ...675.1213R}. Surprisingly, that paper found an intrinsic oxygen abundance scatter of 0.11 dex at a given radius, implying the the ISM is not well mixed. As a consequence, the abundance gradient is difficult to determine if the number of analyzed objects is not large enough (e.g. the slope changes from $-0.027$ to $-0.054$ dex/kpc following \cite{2008ApJ...675.1213R} or \cite{2007MmSAI..78..753M}, with 61 and 14 \HII\ regions, respectively). The only comprehensive study of the metallicity gradient using massive stars (BSGs) to date is that of \cite{2005ApJ...635..311U}, who obtain -0.06 dex/kpc, but only considering 11 stars. M33 was also nearly fully imaged at HST resolution by PHATTER \cite{Willetal21}, thus allowing us to already have a good selection of young stellar clusters to derive the IMF with a 30~m telescope (e.g. lower right panel of Fig.~\ref{M33}). The galaxy hosts NGC~604, the prototype Scaled OB Association, in its outskirts (\cite{Maizetal04a} and upper right panel of Fig.~\ref{M33}). Together with 30~Dor (which has a higher proportion of younger stars) it is the richest low-extinction site of star-formation in the Local Group and, hence, a prime target for a 30~m telescope. In particular, the relatively strong gradient in M33 brings its metallicity from solar at the center to values intermediate between the LMC and the SMC in the outskirts. This makes M33 a perfect laboratory where to independently test our current massive stars winds and evolution knowledge as well as the impact of metallicity and environmental effects of massive star formation and evolution. In parallel, achieving a 2D map of abundances of the blue massive star population of M33 will allow to perform a comprehensive study of both radial and azimuthal variations of metallicity across the disc of M33, providing useful constraints for studies of the chemical evolution of this large-scale pure-disc spiral galaxy.   

\section{Beyond the Local Group} 

$\,\!$\indent Going beyond the Local Group implies losing spatial resolution and being able to analyse a smaller range of stellar masses in the IMF. On the other hand, it provides us with a larger sample of spiral galaxies and, more specifically, of massive face-on grand-design ones. It is there where a larger range of star-formation intensities can be explored and, with that, their effect on the environment. In this respect, the three best galaxies in the local universe are M101, M51, and NGC~6946, only accessible from the northern hemisphere.

\begin{figure}[t]
 \centerline{\includegraphics[width=0.49\linewidth]{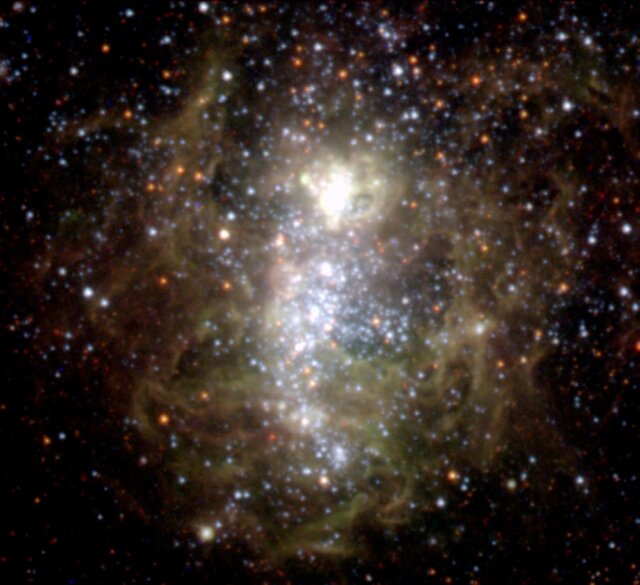} \
             \includegraphics[width=0.49\linewidth]{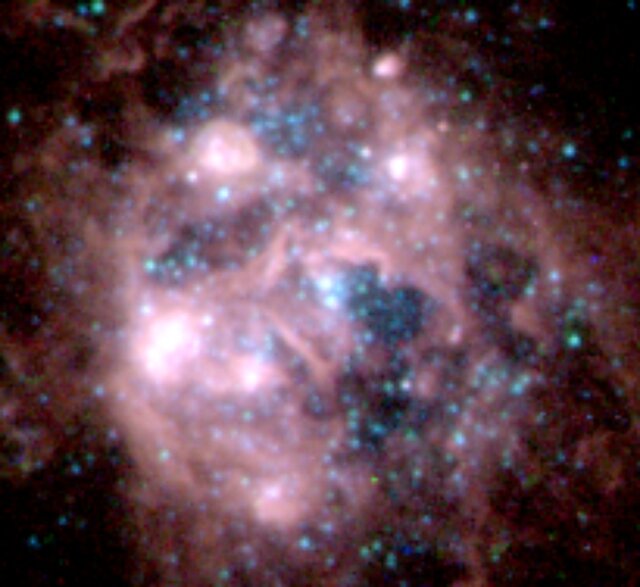}}
 \caption{Two giant H\,{\sc ii} regions in M101 \cite{Bresetal20}. (left) $20.9''\times 19.1''$ (690~pc~$\times$~630~pc) HST/WFC3 mosaic of NGC~5455 with
          F600LP in the red channel, F555W in the green channel, and F438W in the blue channel. (right) $20.9''\times 19.1''$ (690~pc~$\times$~630~pc) 
          HST/WFPC2 mosaic of NGC~5471 with F656N in the red channel, F675W in the green channel, and F547W in the blue channel.}
 \label{M101}
\end{figure}

M101 is a face-on grand-design spiral galaxy located at a distance of 7.5 Mpc. It was other spiral galaxy used by \cite{1971ApJ...168..327S,1975ApJ...199..591S} to establish the presence of an abundance gradient in spirals using \HII\ regions. The most comprehensive determination of the abundance gradient was done by \cite{2003ApJ...591..801K}, who used 20 \HII\ regions to determine a value of $-0.027$~dex/kpc. The study was extended to the inner high-$Z$ disk by \cite{2007ApJ...656..186B}. A recent study by \cite{2025ApJ...991..151B} analysed 13 BSGs in M101, ranging in metallicity from 1.3 Z$_\odot$ to 0.3 Z$_\odot$. Note, however, than the outskirts of M101 have an even lower metallicity, such as the one for NGC~5471 (Fig.~\ref{M101}).

\begin{figure}[t]
 \centerline{\includegraphics[width=\linewidth]{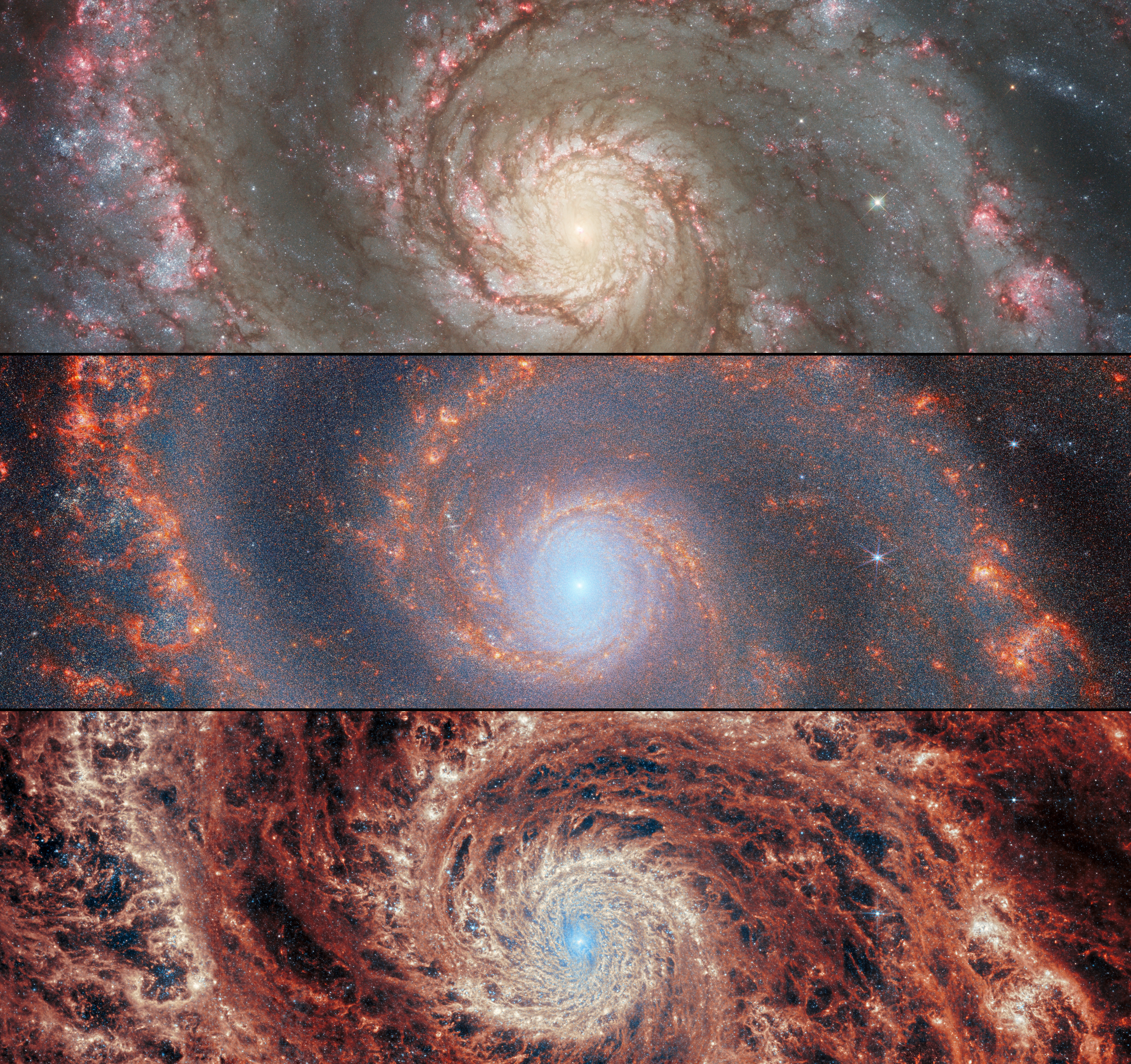}}
 \caption{(left) $5.52'\times 1.71'$ (12.0~kpc~$\times$~3.8~kpc) images of M51 in the [top] visible (HST/ACS-WFC), [center] NIR (JWST/NIRCam), and [bottom] MIR (JWST/MIRI). North is 44$^{\rm o}$ left of the vertical. The top panel can be used to trace low-extinction massive stars and their associated \HII\ regions, as well as the location of high-extinction regions marked by dust lanes. The center panel traces both low- and high-extinction \HII\ regions as well as the red-giant population. The bottom panel traces the warm dust heated by hot massive stars located mostly around dense gas, when looked as a positive image, and the cavities produced by massive-star feedback, when looked as a negative image.}
 \label{M51}
\end{figure}

M51 is another large-design face-on spiral galaxy at a similar distance (7.2~Mpc) that has been interacting with the close-by small galaxy NGC 5195 in the past hundreds of millions of years. This interaction has been certainty affecting its star formation mode, hence making M51 an ideal and unique laboratory for environmental studies (e.g. \cite{2025MNRAS.543.1410J}). Figure~\ref{M51} shows how the ISM ``in negative'' in the form of dark dust lanes in the optical becomes an ISM ``in positive'' in the MIR due to dust emission. 

As a third case, NGC~6946 is also grand spiral with a high inclination at a similar distance as than M101 and M51 (7.7~Mpc). It has a high star-formation rate and a large number of \HII\ regions across its face, that allows for the determination of its 2D abundance distribution \cite{2012A&A...545A..43C}. In particular, \cite{2025MNRAS.539..755B} have detected azimuthal metallicity variations across spiral arms.

%Other spirals: M81, NGC 2403, IC 342, M94 (face-on AGN; DM=28.5)

The massive star population of these galaxies (and some other, such as M81, NGC~2403, IC342 and M94) offers the possibility to extend the studies in M31 and M33. There are only a handful of studies analyzing individual massive stars beyond the Local Group (e.g. \cite{2005ApJ...622..862U} for NGC~300, \cite{2012ApJ...747...15K} for M81 or \cite{2022ApJ...940...32B} for NGC~2403). Thus we could study the upper end of the IMF, the $Z$-dependence of the massive stellar physics (also taking into account the possibility of finding supersolar metallicities in these galaxies). Although spatial resolution will often be insufficient to resolve the stellar clusters with a 30~m-class telescope, we will be able to conduct spectral synthesis analyses in relatively small regions. These could be later extended to further galaxies.  The comparison of results for \HII\ regions with the stellar abundances will provide a solid ground to determine metallicities at large distances. 

\section*{References}

\begin{adjmulticols}{2}{-5mm}{-5mm}
\bibliographystyle{plainurljma}
\bibliography{general}
\end{adjmulticols}

\vfill

\centerline{Text in red in this document corresponds to \href{https://en.wikipedia.org/wiki/Hyperlink}{hyperlinks}.}

\newpage 

\section*{Figures}

%\begin{figure}[t]
% \centerline{\includegraphics[width=\linewidth]{IMG_6700.jpeg}}
% \caption{TBD.}
% \label{TBD}
%\end{figure}

\end{document}